\documentclass[prd,superscriptaddress,twocolumn,nofootinbib,%
  showpacs,amsmath,amssymb,aps,10pt]{revtex4-1}

\usepackage{graphicx}
\usepackage[breaklinks, colorlinks=true]{hyperref}

\newcommand{\CP}[1]{{\mathbb{C}P^{#1}}}
\newcommand{\EntF}{S_{\rm DFT}}
\newcommand{\EntS}{S_{\rm SVD}}
\newcommand{\fF}{f_{\rm DFT}}
\newcommand{\fS}{f_{\rm SVD}}
\newcommand{\betath}{\beta_{\rm th}}
\newcommand{\calD}{\mathcal{D}}
\renewcommand{\Re}{\mathrm{Re}\,}
\renewcommand{\Im}{\mathrm{Im}\,}

\begin{document}
\title{Image-processing the topological charge density
  in the $\CP{N-1}$ model}
\author{Yuya Abe}
\affiliation{Department of Physics, The University of Tokyo, %
  7-3-1 Hongo, Bunkyo-ku, Tokyo 113-0033, Japan}
\author{Kenji Fukushima}
\affiliation{Department of Physics, The University of Tokyo, %
  7-3-1 Hongo, Bunkyo-ku, Tokyo 113-0033, Japan}
\author{Yoshimasa Hidaka}
\affiliation{Nishina Center, RIKEN, %
  2-1 Hirosawa, Wako, Saitama 351-0198, Japan}
\affiliation{iTHEMS Program, RIKEN, %
  2-1 Hirosawa, Wako, Saitama 351-0198, Japan}
\author{Hiroaki Matsueda}
\affiliation{Sendai National College of Technology, %
  Sendai 989-3128, Japan}
\author{Koichi Murase}
\affiliation{Department of Physics, The University of Tokyo, %
  7-3-1 Hongo, Bunkyo-ku, Tokyo 113-0033, Japan}
\author{Shoichi Sasaki}
\affiliation{Department of Physics, Tohoku University, %
  Sendai 980-8578, Japan}

\begin{abstract}
  We study the topological charge density distribution using the
  two-dimensional $\CP{N-1}$ model.  We numerically compute not only
  the topological susceptibility, which is a spatially global quantity
  to probe topological properties of the whole system, but also the
  topological charge correlator with finite momentum.  We perform
  Fourier power spectrum analysis for the topological charge density
  for various values of the inverse temperature $\beta$.  We propose
  to utilize the Fourier entropy as a convenient measure to
  characterize spatial distribution patterns and demonstrate that the
  Fourier entropy exhibits nontrivial temperature dependence.  We also
  consider the snapshot entropy defined with the singular value
  decomposition, which also turns out to behave nonmonotonically with
  the temperature.  We give a possible interpretation suggested from
  the strong-coupling analysis.
\end{abstract}
%\pacs{xxxxx}
\maketitle

%%%%%%%%%%   Introduction   %%%%%%%%%%
\section{Introduction}

Gauge topology is a fundamental aspect of modern field theory.  It
would be an ideal setup for theoretical investigations if a system is
simple but still nontrivial enough to accommodate nonvanishing
topological winding.  The two-dimensional $\CP{N-1}$ model is one of
such ideal theoretical laboratories.  The $\CP{N-1}$ model has the
asymptotic freedom, the linear confining potential, and
instantons~\cite{DAdda:1978vbw,DiVecchia:1981eh,Duane:1981xy,Kunz:1989sh}.
In fact, even on the analytical level, the dynamical mass generation
and the linear confining  potential have been derived in the large-$N$
expansion~\cite{Witten:1978bc,Campostrini:1991tw} as well as in the
strong coupling expansion~\cite{Samuel:1982qm,Plefka:1996ks}.

The $\CP{N-1}$ model has a wide range of applications.
Recently it is a hot and growing area to investigate ``resurgence''
using the $\CP{N-1}$ model, that is, perturbative expandability and
cancellation of ambiguity in each topological sector are intensely
studied~\cite{Dunne:2012ae,Misumi:2014jua,Fujimori:2016ljw}.  Also, we
note that the $\CP{N-1}$ model has plenty of connections to condensed
matter physics systems.  There are several equivalent formulations of
the $\CP{N-1}$ model such as SU($N$) Heisenberg ferromagnets, the
tensor network with three-dimensional loop model which is generalized
to the form of the tensor renormalization group, and so
on~\cite{Kataoka:2010sh,Nahum:2011zd,Nahum:2013qha,Kawauchi:2016dcg,Kawauchi:2016xng,Takashima:2005qh,Roy:2015ars}.
In particular, in Heisenberg spin systems, the $\CP{N-1}$ model is
relevant for a critical phase between the valence-bond solid and the
antiferromagnetic phase~\cite{Motrunich:2008rs,Sandvik:2010ag}.
Furthermore, we point out that recent developments of condensed matter
physics experiments has enabled us to emulate the $\CP{N-1}$ model and
its variants on the optical lattice~\cite{Laflamme:2015wma}.  In this
way, although the $\CP{N-1}$ is a relatively simple and
well-established model, many interesting studies are ongoing to the
present date.

From the point of view of gauge topology, which is of our present
interest, the $\CP{N-1}$ model has a prominent feature that the
topological charge can be defined geometrically~\cite{Berg:1981er} and
it rigorously takes integer values even on the lattice.  There are
several subtle points, however.  Even with rigorous quantization, the
physical interpretation in terms of instantons may become unclear in
some parameter regions.  When the temperature or the coupling constant
is far away from the region corresponding to the continuum limit, the
physical lattice spacing would be too coarse to hold the instantons on
the lattice.  For further discussions on lattice artifacts such as
finite size effects and the critical slowing down of topological
sectors, see
Refs.~\cite{Rossi:1993nz,Irving:1992cr,Rossi:1993nz,DelDebbio:2004xh} 
for example.  Also, in the region with small $N$, quantum fluctuations
would melt instantons by blurring them into quasi-particles.
Therefore, we can perform the instanton gas analysis at large $N$,
while melting instantons causes the precocious scaling in small-$N$
regions~\cite{Andersen:2006sf,Lian:2006ky,Diakonov:1999ae,Maul:2000wb}.
Nevertheless, regardless of the physical interpretation, the
topological winding anyway exists in theory for any parameters, and it
would be desirable to perform analysis on the winding not necessarily
in an intuitive picture of instantons.

The $\CP{N-1}$ model is an intriguing lattice model on its own, and
additionally, one of important model usages is to adopt it as a toy
model for more complicated theories such as QCD-like theories, where
QCD stands for quantum chromodynamics.
Historically speaking, the instantons and the $\theta$-vacuum were
first revealed for QCD in the context to understand color confinement
and spontaneous and anomalous chiral symmetry breaking.  It has been a
widely accepted idea that the chiral condensate that spontaneously
breaks chiral symmetry is induced by fermionic zeromodes associated
with the instanton gauge
background~\cite{Diakonov:1984vw} (see Ref.~\cite{Schafer:1996wv} for
a comprehensive review).  Also, recently, there are significant
progresses in theoretical studies of color confinement based on the
instanton with nontrivial holonomy (see, for example, a proceedings
article in Ref.~\cite{Shuryak:2016vow} and references therein, and
also Ref.~\cite{Fukushima:2017csk} for a recent review).

Now, it is still a challenging problem how to access topologically
nontrivial sectors generally in field theories.  This is the case
also for simple models such as the $\CP{N-1}$ model.  For this
purpose to study topological contents, the topological susceptibility,
$\chi_t$, is the most common observable to quantify fluctuations with
respect to the topological charge.
Roughly speaking, if $\chi_t$ is large, the system should accommodate
more instantons which enhance nonperturbative effects.  There are a
countless number of precedent works on physics implications of
$\chi_t$ in various theories including QCD
(see Refs.~\cite{Borsanyi:2015cka,Kitano:2015fla} for
recent attempts motivated with axion cosmology and
Ref.~\cite{Marsh:2015xka} for a review).  For more phenomenological
applications of the topological susceptibility and related
observables in hadron physics, see a lecture note in
Ref.~\cite{Shore:2007yn}.

Since topological excitations are inherently nonperturbative, the
information available from analytical considerations is limited, and
the numerical Monte-Carlo simulation on the lattice is the most
powerful approach~\cite{Alles:1996nm,*Alles:1997sy,*Alles:2004vi}.
For the $\CP{N-1}$ model, there are many lattice
studies~\cite{Duane:1981xy,Kunz:1989sh,Campostrini:1992ar,Hasenbusch:1993na}.
We shall make a remark that the $\CP{N-1}$ model action has a special
analytical structure such that the sign problem is tamed even 
with a nonzero chemical
potential~\cite{Rindlisbacher:2016cpj,Evans:2016iim} or a $\theta$
term~\cite{Burkhalter:2001hu,Azcoiti:2003qe,Beard:2004jr,Imachi:2004bf,Kawauchi:2016xng},
which is an advantage over other complicated systems.

In this paper we revisit the topological properties in the $\CP{N-1}$
model.  For actual procedures we will regard the two-dimensional field
configurations in the $\CP{N-1}$ model as ``image'' data and will
``image-process'' the distribution of the topological charge density
at various temperatures.  Our motivation partly comes from an analogy
to finite-temperature QCD in which $\chi_t$ changes with temperature.
Actually, we do not restrict our $\CP{N-1}$ model temperature to the
scaling region near the continuum limit;  the $\CP{N-1}$ model is
anyway a well-defined lattice theory for any temperature.
(We will never discuss practical applicability to QCD and the
continuum limit in the present study.)

As a first step, we perform the Fourier power spectrum analysis of the
topological charge density.  This power spectrum amounts to a
momentum dependent generalization of the topological susceptibility,
which has been discussed and partially measured also in QCD in
Refs.~\cite{Shore:1998dm,Fukushima:2001ut,Koma:2010vx}.  The entire
structures of the correlator in momentum space are quite informative,
as we will reveal in this work, but such calculations are
too much resource consuming.  Therefore, it would be much more
convenient if there is any single measure extracting essential
features of the topological charge density distribution as a function
of temperature.  We will propose to use an entropy defined with the
Fourier spectrum.  Also, another observable is obtained to make use of
the singular value decomposition (SVD) of the image data.  As a
standard image-processing tool, the SVD is commonly employed;  see
Refs.~\cite{Matsueda:2012zz,Matsueda:2014rta,Lee:2014ama} for physics
applications of the SVD to image-process the spin configurations.  In
the same way as in the Fourier power spectrum analysis, interestingly,
an entropy is constructed with the SVD eigenvalues, which will be
referred to as the ``snapshot entropy,'' which will turn out to have
nontrivial temperature dependence.

This paper is organized as follows.  In Sec.~\ref{sec:formulation} we 
will review the $\CP{N-1}$ model and the simulation algorithm.  We
will introduce physical observables including the Fourier and the
snapshot entropies in Secs.~\ref{sec:observables} and
\ref{sec:entropies}.  Section~\ref{sec:checks} is devoted to the
numerical setups and the consistency checks with the precedent lattice
simulation results.  In Sec.~\ref{sec:processing} we will report our
main numerical results, namely, the Fourier power spectrum analysis of
the topological charge correlator in Sec.~\ref{sec:power}, the
Fourier and the snapshot entropies in Secs.~\ref{sec:EntF} and
\ref{sec:EntS}, and their correlations in Sec.~\ref{sec:correlation}.
For a possible interpretation of transitional behavior we will discuss
phase boundaries between confinement and infinitesimal deconfinement
from the strong-coupling character expansion.  Finally, in
Sec.~\ref{sec:conclusions}, we will make conclusions and outlook.

%%%%%%%%%%   Formulation   %%%%%%%%%%
\section{Formulation}
\label{sec:formulation}

We will make a brief overview of the numerical lattice simulation in
the $\CP{N-1}$ model.  We adopt the method formulated in
Ref.~\cite{Campostrini:1992ar} to compute physical observables.  We
will also give an explanation of unconventional observables called the
Fourier entropy and the snapshot entropy.

%%%%%%%%%%   $\CP{N-1}$ Model   %%%%%%%%%%
\subsection{$\CP{N-1}$ Model and the Monte-Carlo Method}

The $\CP{N-1}$ model is defined by the following partition function,
\begin{equation}
  Z = \int \calD z\, \calD\lambda\,
  \prod_x \delta (|z(x)|^2-1)\, e^{-\beta H}
  \label{eq:partitionZ}
\end{equation}
with the Hamiltonian,
\begin{equation}
  H = -N\sum_{x,\mu} \bigl[ \bar{z}(x+\hat{\mu})z(x)\lambda_\mu(x)
      + \text{(c.c.)} - 2 \bigr]\,,
  \label{eq:hamiltonian}
\end{equation}
where $z(x)$ represents complex scalar fields with $N$ components
which are constrained as $|z(x)|^2=1$, and (c.c.) is the complex
conjugate of the first term.  The symbol $\hat{\mu}$ denotes the unit
lattice vector.  Since there is no kinetic term, $\lambda_\mu(x)$ is
an auxiliary U(1) link variable.  We will refer to $\beta$ as the
(inverse) temperature throughout this paper.  In some literature
$g\equiv 1/(N\beta)$ is often introduced as a ``coupling constant''
but we will consistently use the inverse temperature $\beta$ only.

The Monte-Carlo simulation consists of two procedures called
``update'' and ``step'' which are explained respectively below.  For
one update we randomly choose a point $x$, and calculate new $z(x)$
and $\lambda_\mu(x)$ at this point $x$ according to the probability
distributions, which will be explicitly given in the subsequent
paragraphs.  One step has $L^2$ updates, where $L$ is the lattice
size.

To make this paper self-contained, let us elucidate how to update
$z(x)$ at a chosen $x$.  We can update $\lambda_\mu(x)$ similarly.
According to a method called the ``over--heat bath method'' in
Ref.~\cite{Campostrini:1992ar} we successively update the
configurations.  It is important that we can write the
$z(x)$-dependent piece of the Hamiltonian as
\begin{equation}
  H = -N \langle z(x), F_z(x)\rangle + \cdots \,,
\end{equation}
where the ellipsis represents terms not involving $z(x)$.  In the
above we introduced the inner product of the $N$ component scalar
fields defined by
$\langle a,b \rangle \equiv \Re \sum_i \bar{a}_i b_i$.  We can easily
infer
\begin{equation}
  F_z(x) = \sum_\mu \bigl[ z(x-\hat{\mu}) \lambda_\mu(x-\hat{\mu})
  + z(x+\hat{\mu}) \bar{\lambda}_\mu(x) \bigr]
\end{equation}
from the Hamiltonian.  The inner product of the $N$ component complex
scalars can be regarded as that of the $2N$ component real scalars,
i.e.,
$\langle a,b\rangle = \sum_i (\Re a_i \Re b_i + \Im a_i \Im b_i)$.
Then, for this real inner product, we can define the relative angle
$\theta$ as
\begin{equation}
  \langle z(x), F_z(x)\rangle = |F_z(x)| \cos\theta(x)\,.
  \label{eq:localHz2}
\end{equation}
Here we used $|z(x)|=1$.  Because the above form depends only on the
relative angle $\theta(x)$, the choice of $z(x)$ is not yet unique for
the same $\langle z(x), F_z(x)\rangle$.  The
over--heat bath method fixes new $z(x)$ uniquely in such a way to
disturb the system maximally.

For the update we first sample a new angle
$\theta$ (that is denoted as $\theta^{\rm new}$ below) according to
the probability\footnote{
The efficient procedure to deal with a sharply peaked function of
$\theta$ is the rejection sampling with Lorenzian fitting;
see Ref.~\cite{Campostrini:1992ar} for details.},
\begin{equation}
  d p_N(\theta) \propto d\theta\, (\sin\theta)^{2N-2}
  e^{\beta N|F_z| \cos\theta} \,,
  \label{eq:probtheta}
\end{equation}
where $(\sin\theta)^{2N-2}$ appears from the measure given by the
surface area of sphere in $2N$-dimensional space.

We next specify new $z(x)$ [that is denoted as
$z^{\rm new}(x)$ below] with chosen $\theta^{\rm new}$.  In the
over--heat bath method, we take $z^{\rm new}(x)$ to minimize an
overlap with original $z(x)$, that is,
$\langle z^{\rm new}(x), z(x)\rangle$ is minimized.  This can be
achieved with
\begin{equation}
  z^{\rm new} = \cos\theta^{\rm new} \frac{F_z}{|F_z|}
  -\biggl( z - \cos\theta \frac{F_z}{|F_z|} \biggr)
  \frac{\sin\theta^{\rm new}}{\sin\theta} \,.
  \label{eq:newz}
\end{equation}
After doing this for $z(x)$, for the same $x$, we perform the update,
$\lambda_\mu(x)\,\to\,\lambda_\mu^{\rm new}(x)$, in the same way.

%%%%%%%%%%   Physical Observables   %%%%%%%%%%
\subsection{Physical Observables}
\label{sec:observables}

In this work we measure physical observables in units of the lattice
spacing $a$.  This means that $\beta$-dependence may enter through the
running coupling $\beta(a)$.  If necessary, we could convert observables
in the physical unit with a typical length scale, i.e.\ correlation
length, though we won't.

The energy density is one of the most elementary physical observables,
that is given by
\begin{equation}
  E = \frac{1}{2NV} \langle H \rangle
  \label{eq:actiondensity}
\end{equation}
with the dimensionless volume $V\equiv L^2$.  It is useful to keep
track of $E$ to monitor if the numerical simulation converges
properly.

We shall define the correlation length.  The U(1) invariance implies
that the basic building block of local physical observables should be
the following local operator
(that is, a counterpart of a mesonic state
in lattice QCD language),
%(that is, a counterpart of the plaquette
%in the lattice gauge field theory),
\begin{equation}
  P_{ij}(x) = \bar{z}_i(x)z_j(x)\,.
  \label{eq:matP}
\end{equation}
The two-point correlation of $P_{ij}(x)$ is
\begin{equation}
  G_P(x,y) = \langle{\rm tr} P(x) P(y) \rangle -\frac{1}{N}\,,
  \label{eq:green}
\end{equation}
where the last term $1/N$ subtracts the disconnected part.  In the
perturbative regime near the scaling region, the correlation function
in momentum space is expected to scale as
\begin{equation}
  \tilde{G}_P(k) \sim \frac{Z_P}{\xi_G^{-2}
    + \sum_\mu 4 \sin^2 (k_\mu/2)} \,,
  \label{eq:greeninPsp}
\end{equation}
where $k_\mu$ is discretized as $k_\mu=2\pi n_\mu/L$ with
$n_\mu=0, 1, 2, \dots, L-1$ for the lattice size $L$
(where we use a notation slightly different from
Ref.~\cite{Campostrini:1992ar}).  We note that the above
form assumes the periodic boundary condition at the spatial edges.
Using the smallest nonzero momentum $k_{(1,0)}\equiv (2\pi/L,0)$, we
can solve the correlation length as
\begin{equation}
  \xi_G^2 = \frac{1}{4 \sin^2 (\pi/L)}
  \biggl[ \frac{\tilde{G}_P(0)}{\tilde{G}_P(k_{(1,0)})} -1 \biggr] \,.
  \label{eq:corrL}
\end{equation}
as given in Ref.~\cite{Campostrini:1992ar}.  As we mentioned in the
beginning of this section, dimensionful observables must scale with
physical $\xi_G$ in the scaling region as this is the only scaling
factor.  For example, in the scaling region in the large-$N$
expansion, the analytical behavior is known as~\cite{Campostrini:1991tw}
\begin{equation}
  \frac{\beta^2 \tilde{G}_P(0)}{\xi_G^2}
  = \frac{3}{2\pi} + O( N^{-1} ) \,.
  \label{eq:largeNconst1}
\end{equation}

Our central interest in this work lies in the topological properties
of the theory.  Here, we adopt the following geometrical definition
of the topological charge density,
\begin{equation}
  \begin{split}
    \rho(x) &\equiv \frac{1}{2 \pi} \Bigl\{
    \arg\bigl[ {\rm tr} P(x+\hat{1}+\hat{2}) P(x+\hat{1}) P(x)
      \bigr]\\
    &\qquad + \arg\bigl[ {\rm tr} P(x+\hat{2})
      P(x+\hat{1}+\hat{2}) P(x) \bigr] \Bigr\} \,,
  \end{split}
  \label{eq:TCD}
\end{equation}
where $\arg$ denotes the principal value of the complex argument
within the interval $(-\pi,\pi]$. Then the quantized topological charge is
\begin{equation}
  Q = \sum_x \rho(x) \,.
  \label{eq:numofinst}
\end{equation}
It is mathematically proven that this $Q$ rigorously takes integer
values even on discretized lattice, which is a preferable advantage to
use the $\CP{N-1}$ model.

Using $Q$ as constructed above, we can define the topological
susceptibility as
\begin{equation}
  \chi_t \equiv \sum_x \langle \rho(x) \rho(0) \rangle
  = \frac{1}{V} \bigl( \langle Q^2 \rangle
  -\langle Q \rangle^2 \bigr)\,.
  \label{eq:Tsuscept}
\end{equation}
In the large-$N$ expansion the analytical behavior of the topological
susceptibility is known as well.  That is, in the leading order of
the large-$N$ expansion, the topological susceptibility obtains as
\begin{equation}
  \chi_t = \frac{3m_0^2}{\pi N} + O(N^{-2})\,,
\end{equation}
where $m_0$ is the vacuum expectation value of an auxiliary field
which gives a dynamical mass for $z(x)$.  In terms of the momentum
cutoff $M_{\rm cut}$, it can be expressed as
\begin{equation}
  m_0^2 = M_{\rm cut}^2 e^{-4\pi\beta}\,,
\end{equation}
which leads to the following expression~\cite{DAdda:1978vbw},
\begin{equation}
  \chi_t(\beta) = \frac{3M_{\rm cut}^2}{\pi N} e^{-4\pi \beta}
  + O(N^{-2})
\label{eq:chitbeta}
\end{equation}
as a function of $\beta$ at the one-loop level.  We can also express
this using $\beta$-dependent $\xi_G$ whose leading form
is~\cite{Campostrini:1991tw}
\begin{equation}
  \xi_G^2 = \frac{1}{6 m_0^2} + O(N^{-2})
  = \frac{1}{6M_{\rm cut}^2} e^{4\pi \beta} + O(N^{-2})\,.
\label{eq:xiGbeta}
\end{equation}
The leading behavior of $\chi_t$ is thus characterized as
\begin{equation}
  \chi_t\, \xi_G^2 = \frac{1}{2\pi N} + O(N^{-2})
  \label{eq:largeNconst2}
\end{equation}
in the scaling region, but this scaling relation does not hold for not
large enough $N$ or $\beta$.

%%%%%%%%%%   Fourier and Snapshot Entropies   %%%%%%%%%%
\subsection{Fourier and Snapshot Entropies}
\label{sec:entropies}

In this paper we will pay our special attention to the Fourier entropy
and the snapshot entropy defined by spatial distribution of $\rho(x)$,
which will be useful for our image-processing purpose.

We shall introduce the Fourier entropy as the Shannon entropy using
the Fourier transformed topological charge density,
$\tilde{\rho}(k)$.  The normalization convention of discrete Fourier
transform (DFT) is chosen as
\begin{equation}
  \tilde{\rho}(k) = \frac{1}{\sqrt{V}}
  \sum_{x^\mu} e^{-ik\cdot x} \rho(x)\,,
  \label{eq:DFT}
\end{equation}
where $x^\mu$ runs over $0,\dots,L-1$ in units of $a=1$.  We first
define the normalized Fourier spectrum as
\begin{equation}
  \fF(k) \equiv \frac{|\tilde{\rho}(k)|^2}{\sum_{k'} |\tilde{\rho}(k')|^2}\,,
\end{equation}
and then the Fourier entropy reads,
\begin{equation}
  \EntF \equiv -\sum_{k} \fF(k) \ln \fF(k)\,.
  \label{eq:FourierS}
\end{equation}
The entropy quantifies how the topological charge density distributes
over space.  For $k$-independent constant $\tilde{\rho}$, $\EntF$ is
saturated at $2\ln L$.

Another quantity which reflects the spatial pattern of $\rho(x)$ is
the snapshot entropy defined with the singular values.  Although our
numerical simulation in the present work uses the square lattice, the
procedure is applicable for general rectangular lattices.  On the
lattice the image of the topological charge density is regarded as an
$L\times L$ real-valued matrix, and its SVD for $x=(x_1,x_2)$ is
written as
\begin{equation}
  \rho(x_1,x_2) = \sum_{n=1}^L
  \lambda^{(n)} U^{(n)}_{x_1} V^{(n)}_{x_2} \,.
  \label{eq:SVD}
\end{equation}
Here, $U^{(n)}$ and $V^{(n)}$ are two sets of orthonormal bases in
$L$-dimensional vector space given by diagonalizing the matrix
$\rho^{\dag} \rho$.  Singular values, $\lambda^{(n)}$, sorted in
descending order are given by the square root of the eigenvalue of
$\rho^{\dag} \rho$.  The two-dimensional image of
$U^{(n)}_{x_1} V^{(n)}_{x_2}$ is referred to as the $n$-th SVD layer,
and the weight for each SVD layer is $\lambda^{(n)}$.  Because all
$\lambda^{(n)}$'s are nonnegative by construction, we can define the
snapshot entropy as the Shannon entropy using $\lambda^{(n)}$.  For
this purpose we first normalize the singular values as
\begin{equation}
  \fS(n) \equiv \frac{\lambda^{(n)}}{\sum_{n'} \lambda^{(n')}}\,,
\end{equation}
and then the snapshot entropy reads,
\begin{equation}
  \EntS \equiv -\sum_{n=1}^L
  \fS(n) \ln \fS(n)\,.
  \label{eq:snapshotE}
\end{equation}
The maximum value of $\EntS$ is $\ln L$.

%%%%%%%%%%   Simulation Setups and Consistency Checks   %%%%%%%%%%
\section{Simulation Setups and Consistency Checks}
\label{sec:checks}

We here detail our numerical simulation processes.  We have performed
the calculations for $N=2$, $3$, $10$, $21$ with the $L=32$ lattice
size to see the $N$ dependence, and for $N=10$ with $L=32$, $64$, and
$128$ lattice sizes to see the $L$ dependence.

For each combination of $(N, L)$, we have initialized the
configuration with the random start (i.e., hot start), and we have
confirmed that thermalization is achieved by 2000 Monte-Carlo steps;
we have checked this by comparing results in the hot and the cold
starts.  After thermalization we take 1000 sampling points with an
interval by 100 steps to measure physical observables.  For the error
estimate of physical observables we use the standard Jack-knife method
with $\text{(bin width)}=10$.  We have chosen the bin width and the
interval to suppress the autocorrelation which is checked by the error
estimate of the energy density at several $(N,L)$.  Our simulation
starts with $\beta=0.1$ and we increase $\beta$ by $0.1$ until
$\beta=1.5$, and for each $\beta$ we repeat the above procedures.

We have quantitatively checked the full consistency of our results and
previous results in Ref.~\cite{Campostrini:1992ar} for the energy
density $E$, the correlation length $\xi_G^2$, and the topological
susceptibility $\chi_t$ at several same $(N,L)$, as listed in
Tab.~\ref{tab:compare}.  In this table, our error estimations include 
only the statistical error from the Monte-Carlo simulation.  In
principle, $\xi_G^2$ might contain additional errors from prescription
of subtracting the disconnected part of the correlation function in
Eq.~\eqref{eq:green}.

%---   table   ---%
\begin{table}
  \begin{tabular}{c@{\hspace{0.1em}}c@{\hspace{1em}}c@{\hspace{1em}}c@{\hspace{1em}}c}
    Our Results \\
    \hline
    ($N,\,L$) & $\beta$ & $E$ & $\xi_G^2$ & $\chi_t$ \\
    \hline
    (2,\,36)  & 1.1 & 0.556245(80) & 12.6(1.7) & 0.008083(85) \\
    (10,\,42) & 0.7 & 0.784321(44) & 5.05(55)  & 0.004997(49) \\
    (10,\,30) & 0.8 & 0.666134(52) & 24.26(31) & 0.000827(94) \\
    (10,\,60) & 0.8 & 0.667047(26) & 21.2(1.1) & 0.000999(10) \\
    (21,\,36) & 0.7 & 0.738992(32) & 14.25(20)  & 0.0005513(81) \\
    \hline \\
    Previous Results \\
    \hline
    ($N,\,L$) & $\beta$ & $E$ & $\xi_G^2$ & $\chi_t$ \\
    \hline
    (2,\,36)  & 1.1 & 0.55593(14) & 12.11(48) & --- \\
    (10,\,42) & 0.7 & 0.78402(13) & 5.52(14)  & 0.00505(11) \\
    (10,\,30) & 0.8 & 0.66591(17) & 25.70(40) & --- \\
    (10,\,60) & 0.8 & 0.66701(8)  & 21.80(47) & 0.00101(4) \\
    (21,\,36) & 0.7 & 0.73888(15) & 14.66(22) & --- \\
    \hline
  \end{tabular}
  \caption{Comparison of our results and previous results in
    Ref.~\cite{Campostrini:1992ar} for various $(N,L)$ and $\beta$. In order to establish the consistency, we have collected $2\times 10^4$ statistics with Jack-knife width $10$ for our results.}
  \label{tab:compare}
\end{table}
%---   table   ---%

\begin{figure}
  \includegraphics[width=\columnwidth]{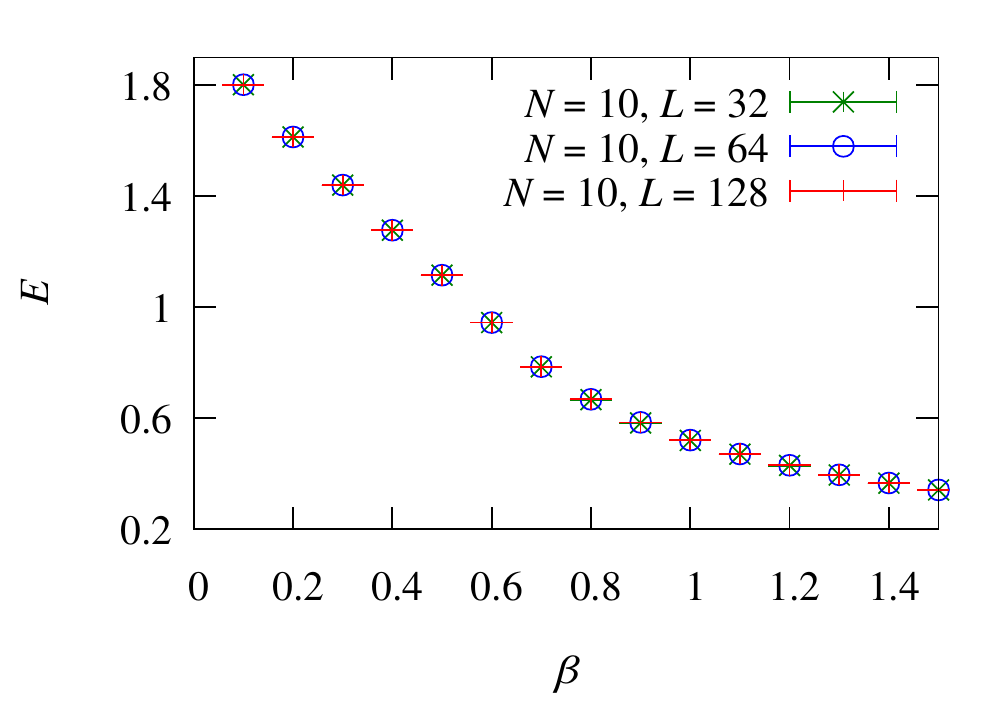}
  \caption{Energy density $E$ as a function of $\beta$ for various
    $L$'s.  The volume dependence is smaller than the dots.}
  \label{fig:EL}
\end{figure}

\begin{figure}
  \includegraphics[width=\columnwidth]{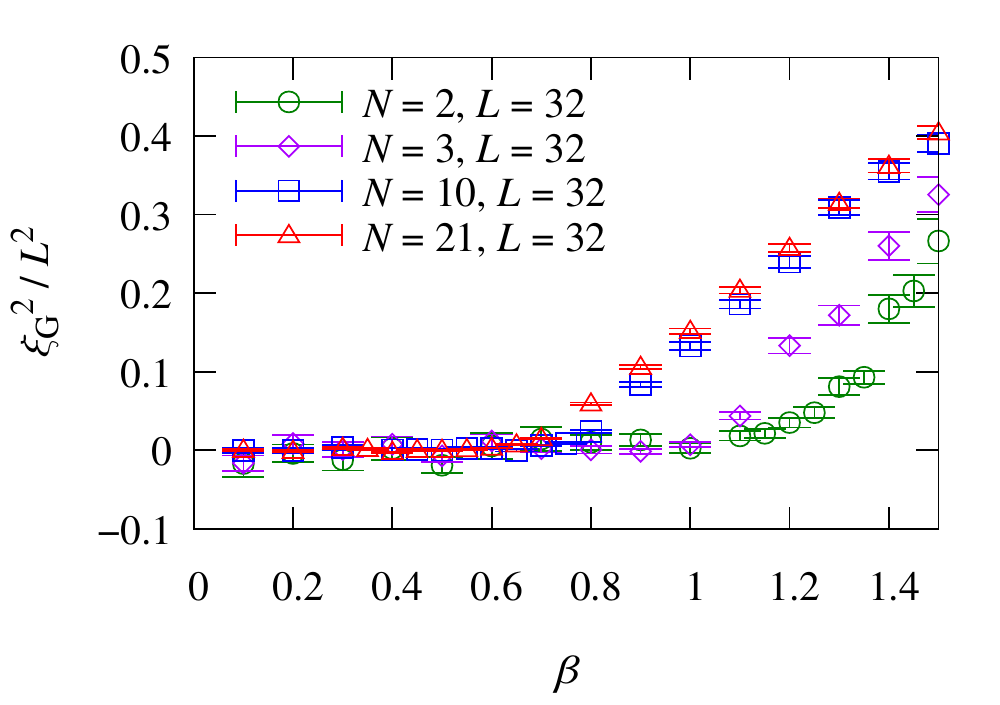}
  \caption{Correlation length squared, $\xi_G^2$, as a function of
    $\beta$ for $L=32$ and various $N$'s.}
  \label{fig:xiG}
\end{figure}

Now, let us proceed to more detailed numerical results for respective
physical observables.  Figure~\ref{fig:EL} shows the energy density
$E$ as a function of $\beta$ for $N=10$ and $L=32$, $64$, $128$.  We
note that this $E$ is a bare one measured in units of the lattice
spacing $a$, and thus a part of $\beta$-dependence is attributed to
$\beta(a)$ as we pointed out.  We make this plot just to see the
volume dependence, and it is clear from Fig.~\ref{fig:EL} that the
volume dependence is negligibly small within the error bars.
Actually, $L=32$ is already sufficiently close to the thermodynamic
limit.  We can understand this from the correlation length shown in
Fig.~\ref{fig:xiG}.  As long as $\beta$ is not too large, the
correlation length is significantly smaller than the lattice size
$L=32$ for any $N$, so that the finite size artifact is expected to be
small already for $L=32$.

%---   figure   ---%
%---   figure   ---$

%---   figure   ---%
%---   figure   ---$

\begin{figure}
  \includegraphics[width=\columnwidth]{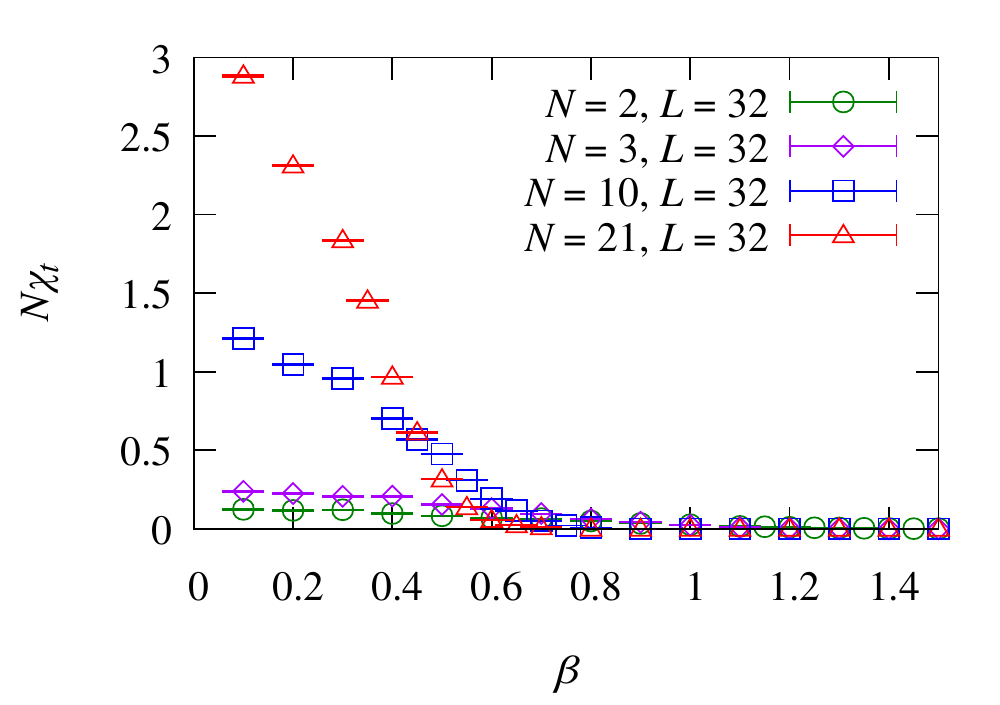}
  \caption{Topological susceptibility, $\chi_t$, as a function of
    $\beta$ for $L=32$ and various $N$'s.}
  \label{fig:chitN}
\end{figure}

In Fig.~\ref{fig:chitN} we show $\chi_t$ as a function of $\beta$ for
$L=32$ and $N=2$, $3$, $10$, $21$.  We see from Fig.~\ref{fig:chitN}
that $\chi_t$ is suppressed for larger $\beta$, which qualitatively
agrees with exponential suppression in Eq.~\eqref{eq:chitbeta}.
We would point out that numerically obtained $N\chi_t$ has quite
nontrivial $N$ dependence, while the convergence of $\xi_G^2$ at large
$N$ is merely monotonic.  In fact, as noticed in Fig.~\ref{fig:xiG},
$\xi_G^2$ at $N=10$ is very close to that at $N=21$, from which one
may want to conclude that $N=10$ could be already a good approximation
for the large-$N$ limit in which analytical formulas are known.  In
Fig.~\ref{fig:chitN}, however, $N\chi_t$ at $N=10$ is not such close
to the $N=21$ results, and furthermore, around $\beta=0.5$, the $N$
dependence is found to be nonmonotonic.  Such prominent differences
between $\xi_G^2$ and $N\chi_t$ clearly indicate that numerically
obtained $N\chi_t$ must have more structures than simple scaling with
$\xi_G^2$.   Section~\ref{sec:processing}  will be devoted to
detailed analysis on this question.

%---   figure   ---%
%---   figure   ---$

%%%%%%%%%%   Finite temperature vs continuum limit   %%%%%%%%%%
\section{Physics Motivation Again}
\label{sec:ftvscl}

Let us rephrase our physics motivation here in a more concrete way in
view of actual data.  As we emphasized the $\CP{N-1}$ model is a
theoretically clean laboratory to study the topologically winding
properties.  Because integer $Q$ is a well-defined quantity for
\textit{any} $\beta$, it is a field-theoretically interesting question
how the topological properties may change with $\beta$, especially for
$\beta$ far from the scaling region.  Indeed, as shown in
Fig.~\ref{fig:chitN}, the topological susceptibility clearly exhibits
non-scaling behavior for $\beta\lesssim 0.7$.

Something nontrivial may be happening there, but the information
available from $\chi_t$ is quite limited.  Non-monotonicity may or may
not be linked to structural change in instanton-like configurations.
In any case there is no way to diagnose what is really happening.

Naively, one could associate $\chi_t(\beta)$ with the typical
instanton size.  For small $\beta$, the correlation length is shorter,
and thus the typical instanton size is expected to be smaller too.
Then, the system with a fixed volume can accommodate more instantons
with smaller size, which enhances the topological susceptibility.  If
this is the case, it would be naturally anticipated that the
topological charge density distribution should contain higher Fourier
components for smaller $\beta$.  The most straightforward strategy in
order to confirm this anticipation is to see the topological charge
density profile in Fourier or momentum space.  Actually, we will
present the Fourier transformed topological charge density in the next
section, which will give us a surprise.

We would make a brief comment here for a possible (in principle, but
unfeasible at the present) application to QCD physics.  We must say
that any QCD discussion itself would be beyond the scope of the
present paper.  Nevertheless, our thinking experiment by means of the
$\CP{N-1}$ model has an intriguing analogy to finite-$T$ QCD physics.
We are discussing $\beta$-dependence of the topological contents, and
it may be sensible to consider $T$-dependence of the topological
contents in QCD;  actually $\chi_t(T)$ has been measured in
lattice-QCD simulations, and then the Fourier transformed $\chi_t$ may
provide us with useful information.

We must not take such an analogy between the $\CP{N-1}$ model and
finite-$T$ QCD, however, for the following reason.  The parameter
$\beta$ in the present study is an inverse temperature for the
classical spin model, but the temperature corresponding to finite-$T$
QCD should be introduced in two-dimensional $\CP{N-1}$ model as a
(1+1)-dimensional quantum field theory on $S^1\times R^1$ where the
period of $S^1$ gives the inverse temperature.  Then, $\beta$ would be
a coupling constant rather than the inverse temperature, whose scaling
property leads to the continuum limit.  Eventually, the continuum
limit of the $\CP{N-1}$ model is to be mapped into a nonlinear sigma
model~\cite{DAdda:1978vbw,Witten:1978bc}.

In contrast to this, the present formulation of the $\CP{N-1}$ model
represents a classical spin system intrinsically defined on the
lattice on $R^2$.  Such a treatment is quite common for condensed
matter physics systems and optical lattice
setups~\cite{Laflamme:2015wma,Azaria:1995wg,Beard:2004jr}.  In this
case the ground state properties, any phase transitions, and
excitation spectra are investigated as functions of $\beta$ including
values far from the scaling region.  There is no imaginary-time
direction corresponding to the finite-temperature quantum field
theory, but two coordinate axes are both spatial.  In this case not
only large-$\beta$ regions but entire $\beta$ dependence is
considered, and the physical unit is provided by the lattice spacing.

%%%%%%%%%%   Image Processing of the Topological Charge Correlator   %%%%%%%%%%
\section{Image Processing of the Topological Charge Correlator}
\label{sec:processing}

So far, we have discussed that our simulation results are fully
consistent with the previous results.  Since the validity has been
confirmed, we are now going into more microscopic views of the
spatial distribution of the topological charge density.  For this
image-processing purpose we perform the Fourier analysis, as
preannounced in the previous section.  As a more convenient measure,
we shall introduce the Fourier entropy and demonstrate its usefulness.
We also discuss a relation to the SVD analysis which is known as a
standard image-processing procedure.

%%%%%   Fourier Spectral Analysis   %%%%%
\subsection{Fourier Spectral Analysis}
\label{sec:power}

\begin{figure*}
  \centering
  \includegraphics[width=\textwidth]{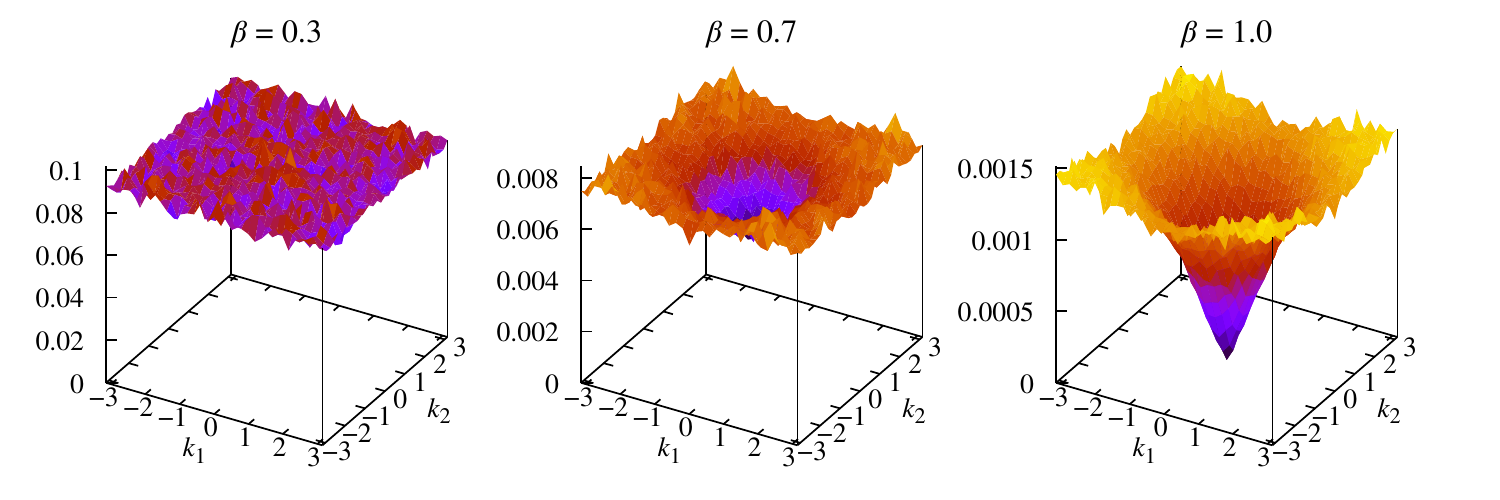}
  \caption{Fourier power spectrum or $\chi_t(k^2)$ for $(N,L)=(10,32)$
    and $\beta=0.3$, $0.7$, $1.0$.  The momenta $k_1$ and $k_2$ run
    from $-\pi$ to $\pi$ in the lattice unit.}
  \label{fig:Fspec_N10L32}
\end{figure*}

In this subsection we present the results from the Fourier spectral
analysis at $(N,L)=(10,32)$ only.  We shall look further into
dependence on $N$ and $L$ when we deal with the entropies in following
subsections.  We already defined the Fourier transformed topological
charge density in Eq.~\eqref{eq:DFT}.  Using this with an ensemble of
1000 configurations, we computed the averaged Fourier power spectrum of
the topological charge density, i.e.,
$\langle|\tilde\rho(k)|^2\rangle$.  We note that $|\tilde{\rho}(k)|^2$
is gauge invariant since $\rho(x)$ is already gauge invariant.
In other words,
$\langle|\tilde{\rho}(k)|^2\rangle$ is nothing but a finite-momentum
extension of the topological susceptibility, i.e.\ $\chi_t(k^2)$ as
defined in Ref.~\cite{Shore:1998dm}.  Hereafter, we shall use a simple
notation of $\chi_t(k^2)$ to mean the Fourier power spectrum.  We
summarize our results for $\chi_t(k^2)$ computations for $\beta=0.3$,
$0.7$, $1.0$ in Fig.~\ref{fig:Fspec_N10L32}.

%---   figure   ---%
%---   figure   ---$

We chose these values of $\beta$ according to qualitative changes in
the Fourier entropy as we will see later.  For $(N,L)=(10,32)$ we will
find that $\beta\approx 0.7$ is a ``threshold'' for suppression of
topological excitations, which we denote by $\betath$.

As long as $\beta \lesssim \betath$, $\tilde{\rho}(k)$ and thus
$\chi_t(k^2)$ spread uniformly over momentum space (see the left panel
in Fig.~\ref{fig:Fspec_N10L32}).  We may say that the topological
charge density is ``white'' then.  In contrast to this, small momentum
components are significantly suppressed for $\beta \gtrsim \betath$ and
we are inclined to interpret this behavior as suppression of
topological excitations.  In fact, it is possible that small momentum
components of $\chi_t(k^2)$ are more diminished with increasing
$\beta$;  the spectral intensity at $k=0$ is nothing but $\chi_t$,
i.e., $\chi_t(0)=\langle|\tilde\rho(0)|^2\rangle = \langle Q^2/V \rangle =
\chi_t$ (where $\langle Q\rangle = 0$), and we already observed
decreasing $\chi_t$ with increasing $\beta$ in Fig.~\ref{fig:chitN}.

It is intriguing that the topological charge correlator at finite
momenta, $\chi_t(k^2\neq 0)$, has such a nontrivial shape in momentum
space even when $\chi_t$ is nearly vanishing.  We can give a
qualitative explanation for this structure. Although the topological
charge itself is robust against perturbative fluctuations, the
correlation function has a nonzero contribution from topologically
trivial sector.  Therefore, we can perform one-loop calculation to
find nonzero $\chi_t(k^2)$ for $k$ large enough to justify
perturbative treatments as
\begin{equation}
  \chi_t(k^2) = \frac{3m_0^2}{\pi N}
  + \frac{3 k^2}{10\pi N} - \frac{k^2}{2(2\pi)^2 \beta N}\,,
\label{eq:perturbative_chi_t}
\end{equation}
where we approximately adopted a continuum theory and we note that the
last term with negative sign would be suppressed for large $\beta$
which is assumed in the continuum limit.  This quadratic rise of
$\chi_t(k^2)$ may partially account for our numerical results at
$\beta > \betath$.
At the same time, however, such an interpretation would invalidate our
n\"{a}ive  anticipation that the system at smaller $\beta$ may
accommodate more instantons with smaller size.  Therefore, we think
that $\chi_t(k^2)$ for large $k$ has inseparable contributions from
perturbative fluctuations and small-sized instantons both.

%%%%%   Fourier Entropy   %%%%%
\subsection{Fourier Entropy}
\label{sec:EntF}

It would be far more convenient if there is, not a whole profile, but
a single observable whose value characterizes changes as in
Fig.~\ref{fig:Fspec_N10L32}.  We propose to use the Fourier entropy
defined in Eq.~\eqref{eq:FourierS} for diagnosis.

%---   figure   ---%
\begin{figure}
  \includegraphics[width=\columnwidth]{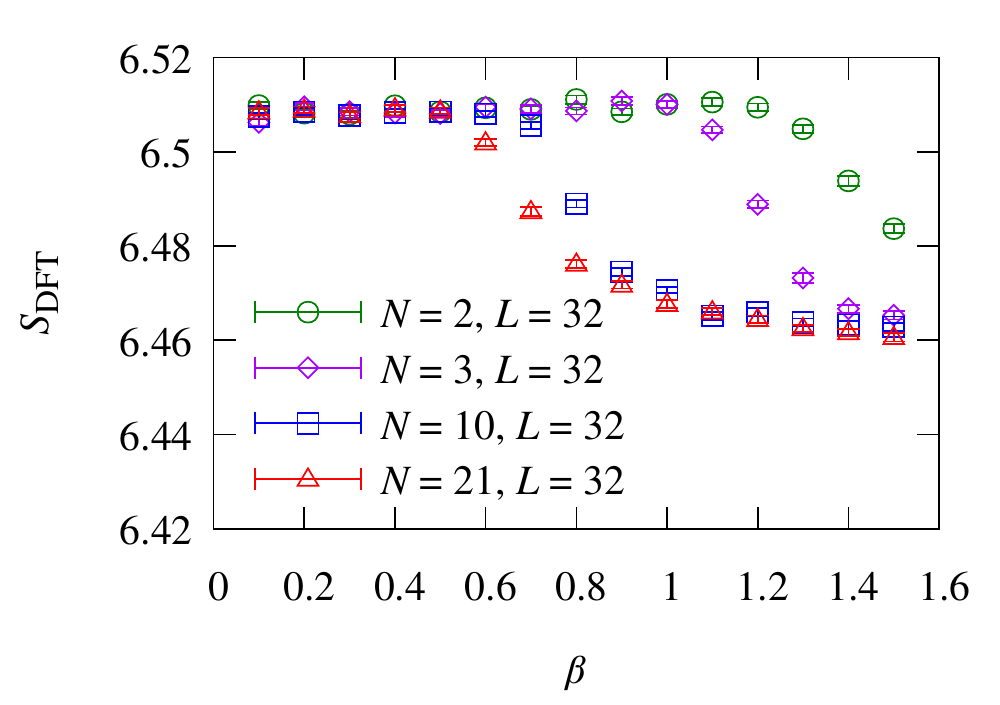}
  \caption{Fourier entropy as a function of $\beta$ for $L=32$ and
    various $N$.}
  \label{fig:EntF-N-dependence}
\end{figure}
%---   figure   ---$

%---   figure   ---$
\begin{figure}
  \centering
  \includegraphics[width=\columnwidth]{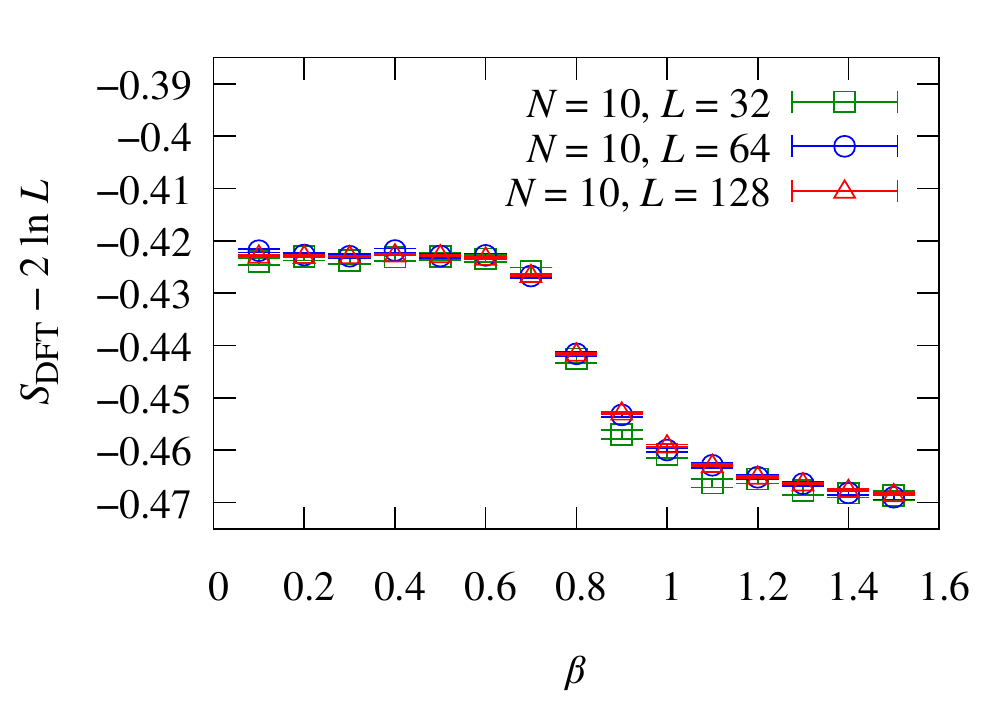}
  \caption{Subtracted Fourier entropy, $\EntF-2\ln L$, as a function
    of $\beta$ for $N=10$ and various $L$.}
  \label{fig:EntF-L-dependence}
\end{figure}
%---   figure   ---$

Figure~\ref{fig:EntF-N-dependence} shows
the Fourier entropy as a function of $\beta$ for $L=32$ and $N=2$,
$3$, $10$, $21$.  We see that the Fourier entropy stays constant for
$\beta \lesssim \betath$, where $\betath$ turns out to depend on $N$.
Then, the Fourier entropy starts dropping at $\betath$.  As compared
to Fig.~\ref{fig:Fspec_N10L32}, the threshold behavior is clearly
manifested and precisely quantified in
Fig.~\ref{fig:EntF-N-dependence}.

We also checked the $L$ dependence of the Fourier entropy.  We note
that the saturated value of $\EntF$ is $2\ln L$ and thus it contains
logarithmic $L$ dependence.  Interestingly, we found that the
subtracted Fourier entropy, $\EntF-2\ln L$, seems to have a
well-defined thermodynamic limit.  That is, we show the subtracted
Fourier entropy as a function of $\beta$ in
Fig.~\ref{fig:EntF-L-dependence} for $N=10$ and $L=32$, $64$, $128$.
The subtracted Fourier entropy barely has $L$ dependence as confirmed
in Fig.~\ref{fig:EntF-L-dependence}.  This result is consistent with
our previous discussion that $L=32$ is already close to the
thermodynamic limit.

From these results we could deduce changes in the topological contents
as follows.  For $\beta \lesssim \betath$ the topological
fluctuations are large as seen in $\chi_t$ but the topological
contents are white.  For $\beta \gtrsim \betath$ where $\EntF$ starts
decreasing, the topological contents with small $k$ are breached,
which implies that large-sized domains (possibly instantons) are
suppressed simultaneously as the correlation length becomes larger.

%%%%%   Snapshot Entropy   %%%%%
\subsection{Snapshot Entropy}
\label{sec:EntS}

The Fourier power spectrum is a useful device for the image
processing, and a more conventional alternative is the SVD analysis
which is suitable for coarse-graining the image.  Let us compute the
snapshot entropy using the SVD and check whether anything nontrivial
appears near $\betath$ or not.

\begin{figure}
  \includegraphics[width=\columnwidth]{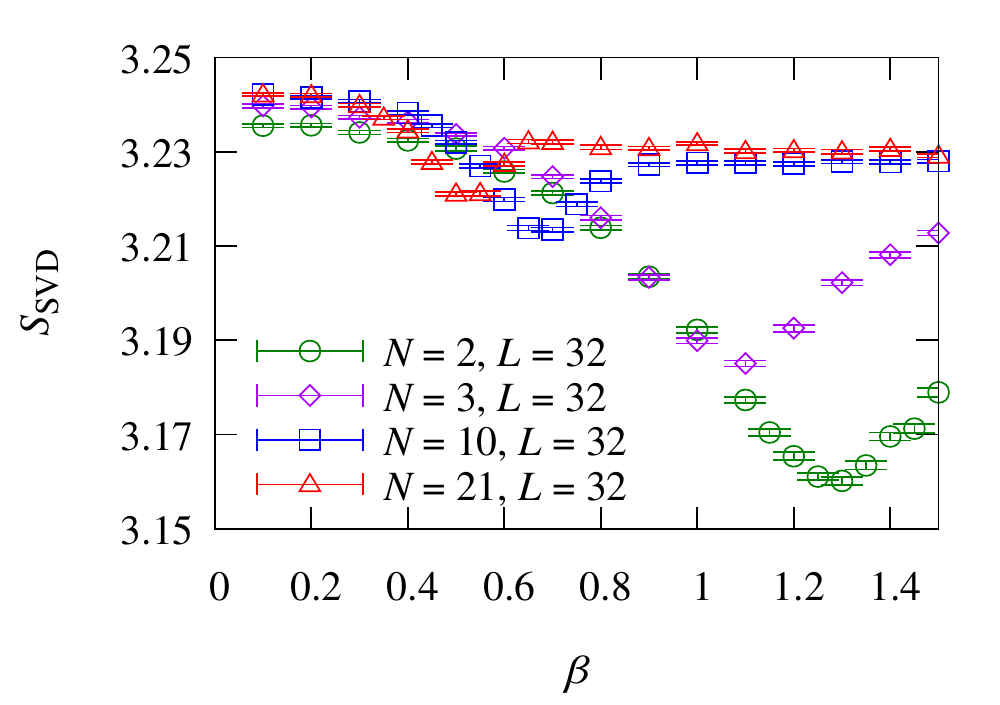}
  \caption{Snapshot entropy, $\EntS$, as a function of $\beta$ for
    $L=32$ and various $N$.}
  \label{fig:SN}
\end{figure}

In Fig.~\ref{fig:SN} we plot $\EntS$ as a function of $\beta$ for
$L=32$ and $N=2$, $3$, $10$, $21$, which is a SVD counterpart of the
previous plot in Fig.~\ref{fig:EntF-N-dependence}.

%---   figure   ---%
%---   figure   ---$

%---   figure   ---%
\begin{figure}
  \includegraphics[width=\columnwidth]{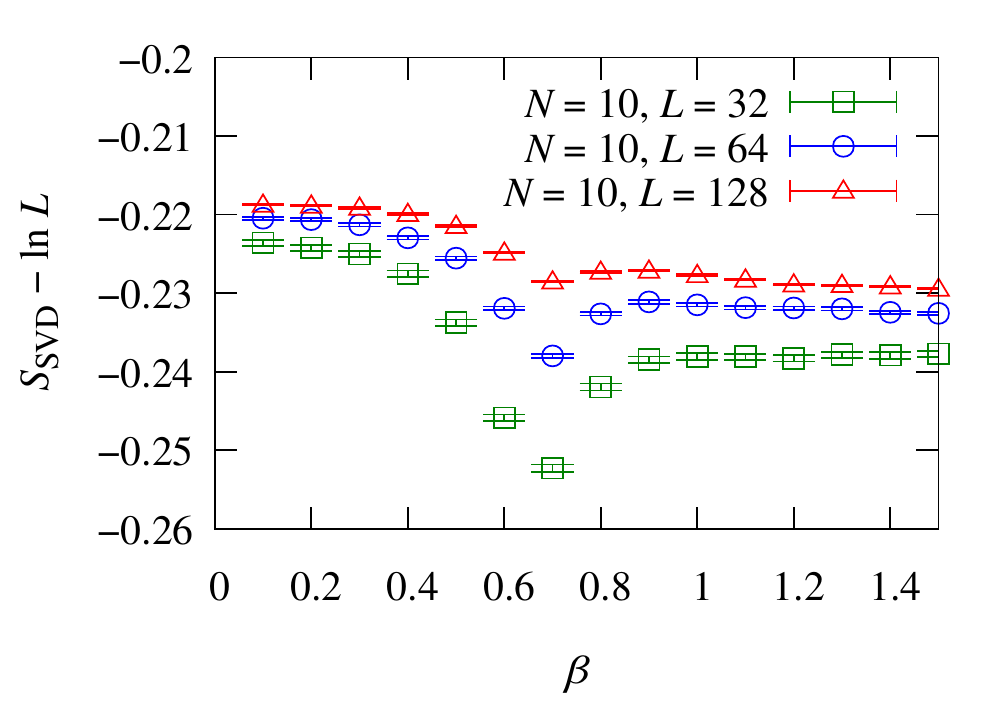}
  \caption{Subtracted snapshot entropy, $\EntS-\ln L$, as a function
    of $\beta$ for $N=10$ and various $L$.}
  \label{fig:SL}
\end{figure}
%---   figure   ---$

Instead of clear threshold behavior in
Fig.~\ref{fig:EntF-N-dependence}, we found that $\EntS$ exhibits a dip
around $\betath$ as seen in Fig.~\ref{fig:SN}.  Similarly to the
Fourier entropy, in the case of $\EntS$, the depth and the location of
the dip depend on $N$.  Interestingly, the $L$ dependence of $\EntS$
is quite different from that of $\EntF$.  Figure~\ref{fig:SL} shows
the subtracted snapshot entropy, $\EntS-\ln L$, for $N=10$ and $L=32$,
$64$, $128$.  We see that some sizable $L$ dependence remains even
after the subtraction, which makes a sharp contrast to
Fig.~\ref{fig:EntF-L-dependence}.  Such $L$ dependent results are
highly nontrivial;  we tested $L$-scaling analysis, but the $L$
dependence seen in Fig.~\ref{fig:SL} turned out not to obey simple
scaling.  We would point out an example of analytically calculable
$\EntS$;  in a real random matrix theory,
$\EntS^{\rm (RM)} = \ln L  - \pi/4$ is known~\cite{Matsueda:2012zz}.
Thus, $L$ dependent $\EntS$ may already indicate that the theory under
consideration has some interesting features.

In order to locate $\betath$ for numerical simulations in a finite
size box, $\EntS$ is as useful as $\EntF$.  However, it is evident
from Fig.~\ref{fig:SL} that the dip depth becomes shallower for larger
$L$, implying that the dip may eventually disappear in the
thermodynamic limit unless we know the proper $L$ scaling.  Therefore,
for a practical usage, $\EntF$ would be a more tractable choice.

%---   figure   ---%
\begin{figure*}
  \centering 
  \includegraphics[width=\textwidth]{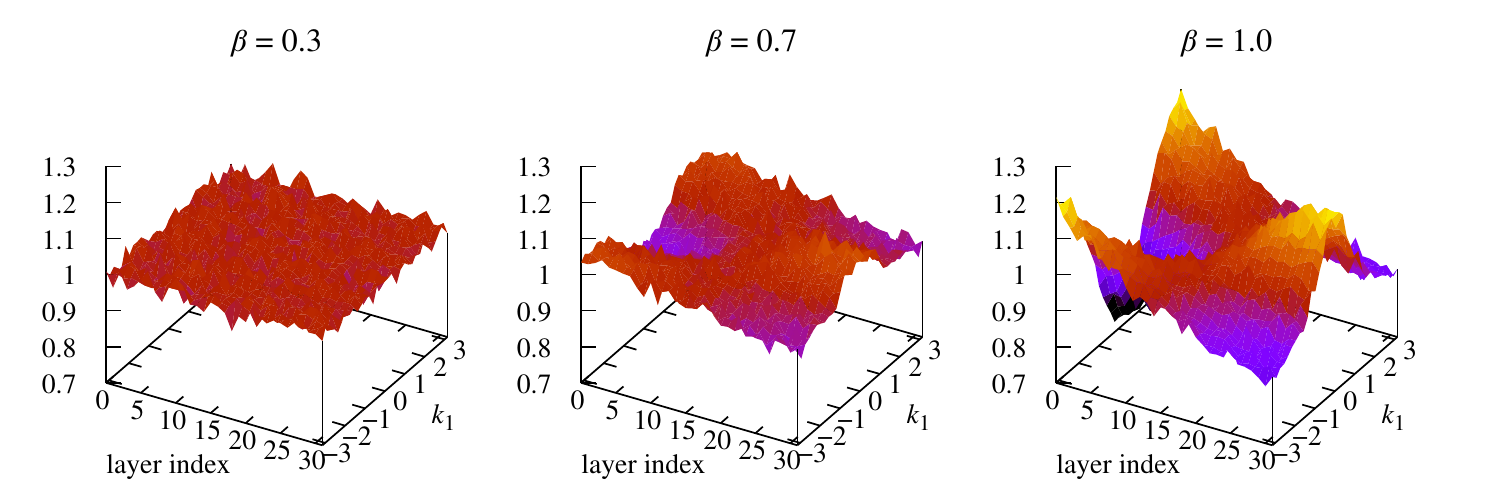}
  \caption{SVD layer dependence of the Fourier spectrum for 
    $(N,L)=(10,32)$ and $\beta=0.3$, $0.7$, $1.0$.}
  \label{fig:SVDFspec_N10L32M32}
\end{figure*}
%---   figure   ---$

%%%%%   Correlation between Fourier and Snapshot Entropies   %%%%%
\subsection{Correlation between Fourier and Snapshot Entropies}
\label{sec:correlation}

It would be instructive to clarify a possible connection of the
Fourier spectrum and the SVD spectrum of the topological charge
density.  To this end we have calculated the Fourier spectrum of each
SVD layer of the topological charge density.  We note that each SVD
layer is a direct product of two vectors, $U^{(n)}_{x_1} V^{(n)}_{x_2}$,
so that the Fourier transform with respect to two spatial directions
is trivially factorized, and we can separately discuss
$\tilde{U}^{(n)}(k_1)$ [that is a Fourier transform of
  $U^{(n)}_{x_1}$] and $\tilde{V}^{(n)}(k_2)$ [that is a Fourier
  transform of $V^{(n)}_{x_2}$].  From symmetry between 1, 2
directions, clearly, it is sufficient to consider only
$\langle|\tilde{U}^{(n)}(k_1)|^2\rangle$ without loss of generality,
and Fig.~\ref{fig:SVDFspec_N10L32M32} shows our results.  We recall
that in our convention a smaller SVD layer index corresponds to a
larger SVD eigenvalue.  We see that the Fourier spectrum is ``white''
at $\beta \lesssim \betath$ for all the SVD layers as is the case in
the left panel of Fig.~\ref{fig:SVDFspec_N10L32M32}.  Some
characteristic patterns start emerging around $\betath\approx 0.7$ as
observed in the middle and the right panels of
Fig.~\ref{fig:SVDFspec_N10L32M32}, namely, images with smaller layer
index (i.e., larger SVD eigenvalue) are more dominated by large
momentum modes, while images with larger layer index contain smaller
momentum modes.  Interestingly, this present situation is quite
unusual;  if the SVD is used for the image processing of ordinary
snapshot photographs, usually, smaller SVD index layers would
typically correspond to a partial image with smaller momenta or larger
spatial domains.  A general trend is that such an ordinary
correspondence in the image processing holds for classical systems,
and for quantum systems the correspondence could be reversed due to
quantum fluctuations.  It is a subtle question whether $\rho(x)$
belongs to classical or quantum class.  Our numerical results suggest
a quantum nature, so that if we want to coarse-grain $\rho(x)$, we
should remove SVD layers from the smallest index.

The physical interpretation of the Fig.~\ref{fig:SVDFspec_N10L32M32} is 
rather straightforward contrary to the image-processing point of view. 
As the $\beta$ grows, larger momentum components become relevant in the 
topological charge density spectrum because of the renormalization scaling 
of the physical length unit.  In the context of the QCD physics, the 
instantons simply melt away at larger $\beta$ (that is, larger
physical temperature).  Or, equivalently, larger $\beta$ prohibits
topological excitations in larger scale if we regard $\CP{N-1}$ model
as the spin model in the condensed matter systems.

%%%%%%%%%% Discussions   %%%%%%%%%%
\section{Discussions from the strong-coupling expansion}
\label{sec:discussions}

We found that the $\CP{N-1}$ model may have transitional behavior as a
function of $\beta$, which is not clear in the topological
susceptibility $\chi_t$ but evidently seen in the Fourier entropy as
well as the snapshot entropy.  We also clarified that this
transitional change is related to the spatial distribution of the
topological charge density.  Then, an interesting question is what the
underlying \textit{physics} should be.

%---   figure   ---%
\begin{figure}
  \centering 
  \includegraphics[width=0.9\columnwidth]{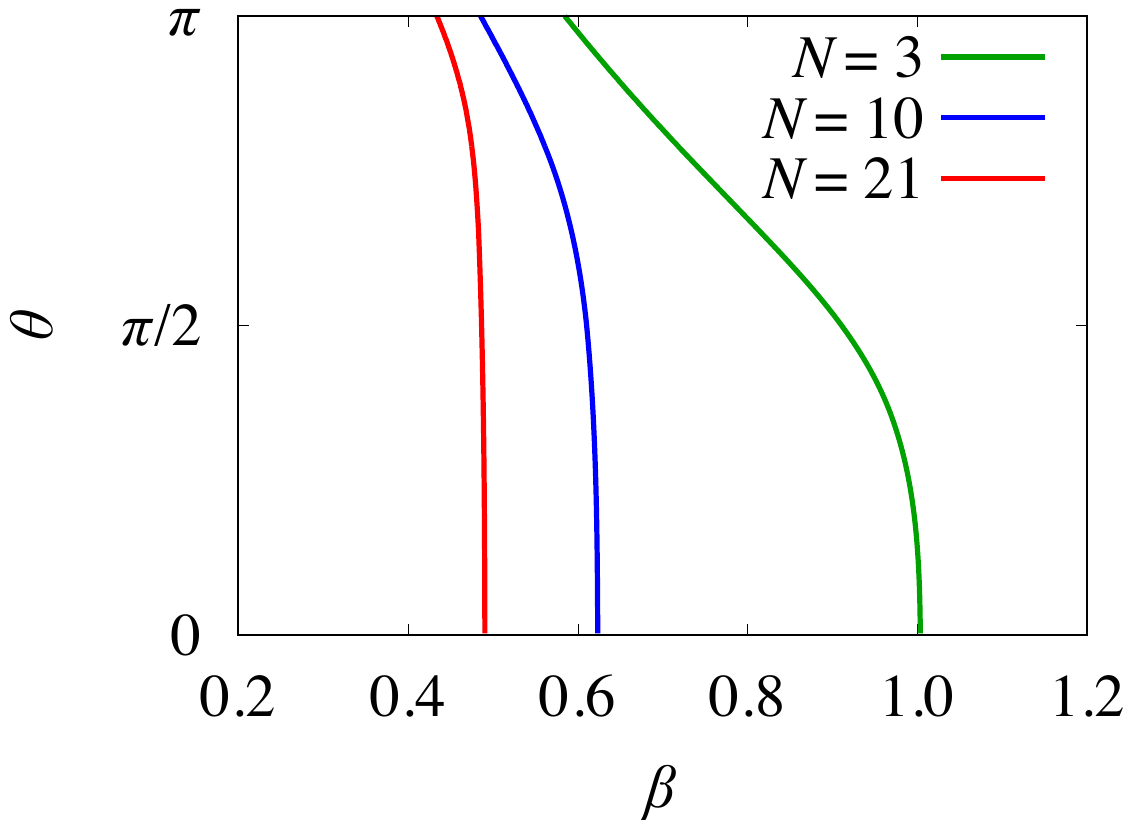}
  \caption{Phase boundary determined from the strong-coupling
    character expansion up to 10 order.  The left (i.e., smaller
    $\beta$) region from the boundary line indicates the full
    confinement phase, while the right (i.e., larger $\beta$) region
    the phase with infinitesimal confinement loss.}
  \label{fig:phase}
\end{figure}
%---   figure   ---$

A hint comes from the analysis in the large-$N$ limit in which a
first-order phase transition occurs with a well-defined
$\beta_{\rm c}$.  As argued in Ref.~\cite{Luscher:1981tq}, the system
is completely disordered for $\beta < \beta_{\rm c}$, while the system
for $\beta > \beta_{\rm c}$ is still disordered but has a finite
correlation length.  These changes are qualitatively consistent with
our results of Fig.~\ref{fig:Fspec_N10L32}.

We can proceed to a further quantitative consideration using the
strong-coupling expansion.  The pressure or the free energy $F$ is
expanded up to order 10 in the character
expansion~\cite{Plefka:1996ks}.  The physics motivation lies in the
phase structure as a function of the coupling $\beta$ and the
topological angle $\theta$.
We note that a finite $\theta$ causes the sign problem
  and the lattice numerical simulations would not work reliably then.
The string tension between charges
$\pm e$ is given by
$\sigma(e,\theta,\beta)=F(\theta+2\pi e,\beta)-F(\theta,\beta)$
which is proportional to $\partial F(\theta,\beta)/\partial\theta$
in the limit of $e\to 0$.  Then, the extremal condition,
$\partial F(\theta,\beta)/\partial\theta=0$, indicates a phase
transition associated with loss of confinement of infinitesimal
charges.  The phase structure inferred from the strong-coupling
expansion is illustrated in Ref.~\cite{Kawauchi:2016dcg} in a way
extrapolated to $\theta\simeq 0$ regions for $N=2$.  Interestingly,
the phase boundary approaches $\beta_{\rm c}=1.2\sim 1.3$ as
$\theta\to 0$ (see Fig.~2 of Ref.~\cite{Kawauchi:2016dcg}).  Of
course, the strong-coupling analysis may lose validity at such large
$\beta$, but the coincidence between this $\beta_{\rm c}=1.2\sim 1.3$
and our $\beta_{\rm th}$ for $N=2$ read from
Figs.~\ref{fig:EntF-N-dependence} and \ref{fig:SN} is quite
suggestive.  Taking this quantitative coincidence seriously, as shown
in Fig.~\ref{fig:phase}, we drew the phase boundaries for other $N$'s
from the condition, $\partial F(\theta,\beta)/\partial\theta=0$, using
the expressions in Ref.~\cite{Plefka:1996ks}.  Then, those phase
boundaries hit $\theta\simeq 0$ at certain values of $\beta$ which are
all very close to our $\beta_{\rm th}$.

We would not claim at present that our $\beta_{\rm th}$ can be
identified as the critical value of $\beta$ corresponding to the
deconfinement phase transition.  It is still under investigations
whether there is a critical $\beta$ at all;
Our present data show no genuine phase transition but
  only a smooth crossover with transitional behavior.
One exotic scenario conjectured in preceding works is that
$\beta_{\rm c}\to \infty$ as $\theta\to 0$, and if so, the strong-$CP$
problem may be understood as a requirement for confinement as argued
in Refs.~\cite{Plefka:1996ks,Kawauchi:2016dcg}.
Our present results do not support $\beta_{\rm c}\to
  \infty$ but not strictly exclude this scenario.
It is still logically possible that, even in such a situation, the phase boundaries
may take steeply bending shapes near $\theta\simeq 0$ and transitional
remnants may be detectable in the simulation at $\theta=0$ like ours.

However, a more straightforward interpretation would be the following:
It is not easy to think of such an abrupt change of $\beta_{\rm c}$ in
the small-$\theta$ region.  In fact, Fig.~\ref{fig:phase} implies
almost no $\theta$-dependence of $\beta_{\rm c}$ for not large values
of $\theta$ (e.g., $\theta \lesssim \pi/4$).  For sufficiently (but
not too) small $\theta$, we see that
$\partial F/\partial\theta\bigr|_{\theta} \simeq
  \partial F/\partial\theta\bigr|_0
  + \partial^2 F/\partial\theta^2\bigr|_0\theta = 0$
leads to $\partial^2 F/\partial\theta^2\bigr|_0 \propto \chi_t \simeq 0$.
Therefore, such a coincidence between $\beta_{\rm c}$ (determined from
$\partial F/\partial\theta= 0$) and $\beta_{\rm th}$ (where
$\chi_t\simeq 0$) is not very surprising.  Moreover, we can give an
intuitive and qualitative argument in favor of
$\beta_{\rm c}\simeq\beta_{\rm th}$.

Physically speaking, the confined phase is known to correspond to
the \textit{disordered state}~\cite{Svetitsky:1985ye}, in which random
distributions in Fourier space should be flat as seen in the left of
Fig.~\ref{fig:Fspec_N10L32} for $\beta<\beta_{\rm th}$.  The point is,
as mentioned in Sec.~\ref{sec:entropies}, the Fourier and the snapshot
entropies take the \textit{maximum} values with uncorrelated random
distributions.  Therefore, if randomness in Fourier space is partially
lost as is the case in the middle of Fig.~\ref{fig:Fspec_N10L32} for
$\beta \simeq \beta_{\rm th}$, the Fourier entropy starts decreasing.
Again, we should remind that it is hard to locate a threshold
precisely for such a smooth crossover, and its precise location would
depend on the choice of observables and prescriptions.

Indeed, how the crossover looks like is quite different for the
snapshot entropy, which starts decreasing around $\beta$ smaller than
$\beta_{\rm th}$.  This is so, as hinted by
Fig.~\ref{fig:SVDFspec_N10L32M32}, since the SVD layers could have
nontrivial structures in Fourier space already when the original
configuration as a superposition of the SVD layers does not yet
develop significant structures.  In this sense, the snapshot entropy
is more sensitive to microscopic substructures.  However, we would
stop such discussions on the snapshot entropy.  As we briefly
mentioned before, unlike the Fourier entropy, the thermodynamic limit
of the snapshot entropy seems to be ill-defined, which itself is quite
surprising.  There might be interesting underlying mechanism, but it
would be out of scope of the current work.  For our claim of sudden
changes of matter around $\beta_{\rm th}$ and a possible
interpretation in terms of infinitesimal deconfinement around
$\beta_{\rm c}$, the Fourier entropy, which looks well-defined in the
thermodynamic limit, should be sufficient.

For more clarification we need to perform simulations at nonzero
$\theta$, but for this purpose, we should circumvent the sign problem
of the Monte-Carlo method.  Apart from the sign problem, once we could
establish that our entropies are sensitive to the phase boundaries, in
turn, our entropies should be quite useful probes to quantify
$\beta_{\rm c}$ precisely.

%%%%%%%%%%   Conclusions   %%%%%%%%%%
\section{Conclusions}
\label{sec:conclusions}

We conducted the numerical Monte-Carlo simulation using the $\CP{N-1}$
model.  We first checked the simulation validity by comparing our
results with the precedent studies for $N=2$, $3$, $10$, and $21$.  We
then scrutinized the spatial distribution of the topological charge
density for various inverse temperature $\beta$ by means of the Fourier
analysis and the singular value decomposition.  There, we found that
the Fourier power spectrum of the topological charge density is rather
structureless or ``white'' in momentum space for small $\beta$, while small
momentum components become diminished for $\beta$ above a certain
threshold $\betath$, which is possibly interpreted as a suppression of
large-sized instantons at large $\beta$.  At the same time, nontrivial
structures (i.e., a drop in the Fourier entropy and a dip in the
snapshot entropy) appear also around $\betath$.  We gave discussions
based on the strong-coupling character expansion in an extended
$\beta$-$\theta$ plane.  Our numerical value of $\betath$ is
suggestively close to critical $\beta_{\rm c}$ of the phase boundary from the
strong-coupling expansion at $\theta\to 0$.  We gave qualitative
arguments in favor of $\beta_{\rm th}\simeq \beta_{\rm c}$ and
explained the decreasing behavior of the Fourier entropy accordingly,
while the snapshot entropy seems to be ill-defined in the
thermodynamic limit and thus its behavior is not under theoretical
control in the current analysis.

Nevertheless, we clarified a correlation between the Fourier and the snapshot
entropies.  In contrast to the ordinary image-processing of picture
images, SVD layers with larger SVD eigenvalues turn out to be
dominated with higher momentum components.  Thus, a cooling method
could be implemented by removing the SVD layers from lower index (with
larger SVD eigenvalue).  A striking finding in this work is that the
Fourier power spectrum of the topological charge density or the finite
momentum extended topological susceptibility, $\chi_t(k^2)$, has
nontrivial momentum dependence even at large $\beta$ when the
topological susceptibility itself is nearly vanishing.  This indicates
that the topological contents of the theory are not necessarily empty
even when the topological susceptibility approaches zero if one
explores finite momentum regions.  One might think that nonzero
topological fluctuations at finite momenta may arise purely from
perturbative loops and may not necessarily involve topological
windings.  This is true, and nevertheless, this situation is still
interesting.  We would recall that such property of $\chi_t(k^2)$ is
reminiscent of the sphaleron rate which is a real-time quantity
analytically continued from the topological susceptibility.  In a
seminal work in Ref.~\cite{Arnold:1987zg} it has been shown that the
sphaleron rate has finite contributions from zero winding sector,
though the analytical continuation makes such terms disappear in the
topological susceptibility.  In other words, the sphaleron rate is a
finite frequency extension from $\chi_t$, and we are discussing a
finite spatial momentum extension from $\chi_t$, and both are as
nontrivial.

One interesting direction would be the lattice calculation of
$\chi_t(k^2)$ in the pure Yang-Mills theory at high temperature where
$\chi_t$ itself is vanishingly small.  Then, a nontrivial question is
whether the pure Yang-Mills theory and QCD may have the behavior of
increasing $\chi_t(k^2) \propto k^2$ or not.
Alternatively, the measurement of the Fourier entropy using the
topological charge density as a function of the physical temperature
should be in principle feasible in the pure Yang-Mills theory.  Our
results imply that the Fourier entropy would show some
nonmonotonic behavior near $T_c$, which could be tested in the pure
Yang-Mills theory in order to quantify how much ``white'' the
topological contents are.  It is tricky, however, how to carry out the SVD
for three or higher dimensional data, and so the applicability of the
snapshot entropy is limited to two-dimensional field theories.

An interesting future problem with special attention to the lattice
$\CP{N-1}$ model would be inclusion of interaction terms with farther
neighborhood, with which nontrivial phase structures are realized.
Another challenging problem in the $\CP{N-1}$ model is how to identify
the instantons and the bions numerically in the $\CP{N-1}$ model
simulations.  In principle, all information on such special
configurations should be encoded on the Fourier power spectrum of the
topological charge density.  Such questions should deserve judicious
investigations in the future.

\acknowledgments
We thank Philippe~de~Forcrand and Massimo~D'Elia for useful
conversations.
This work was supported by Japan Society for the Promotion of Science
(JSPS) KAKENHI Grant No.\ 15H03652, 15K13479, 16K17716, 17H06462,
and 18H01211.

\bibliography{cpN}
\bibliographystyle{apsrev4-1}

\end{document}